%% file: manuscript_arxiv.tex
\documentclass[twocolumn,amsmath]{aastex702}

\shorttitle{A circumbinary planet around KIC~9028474}
\shortauthors{Albrecht et al.}

\graphicspath{{./}{plots/}}

\begin{document}

\title{The BANANA Project: A Super-Jupiter on a 6.7-Year Circumbinary Orbit}

\author[orcid=0000-0003-1762-8235,gname=Simon,sname=Albrecht]{Simon H.\ Albrecht}
\affiliation{Department of Physics and Astronomy, Aarhus University,
	Ny Munkegade 120, 8000 Aarhus C, Denmark}
\email[show]{albrecht@phys.au.dk}
\correspondingauthor{Simon H.\ Albrecht}

\author[orcid=0000-0002-1027-0990,gname=Subo,sname=Dong]{Subo Dong}
\affiliation{Department of Astronomy, School of Physics, Peking University,
	5 Yiheyuan Road, Haidian District, Beijing 100871, People's Republic of China}
\affiliation{Kavli Institute for Astronomy and Astrophysics, Peking University,
	Yi He Yuan Road 5, Hai Dian District, Beijing 100871, China}
\affiliation{National Astronomical Observatories, Chinese Academy of Sciences,
	20A Datun Road, Chaoyang District, Beijing 100101, People's Republic of China}
\email{dongsubo@pku.edu.cn}

\author[orcid=0000-0002-4265-047X,gname=Joshua,sname=Winn]{Joshua N.\ Winn}
\affiliation{Department of Astrophysical Sciences, Princeton University,
	4 Ivy Lane, Princeton, NJ 08544, USA}
\email{jnwinn@princeton.edu}

\author[orcid=0000-0001-7568-6674,gname=Silke,sname=Dainese]{Silke Dainese}
\affiliation{Department of Physics and Astronomy, Aarhus University,
	Ny Munkegade 120, 8000 Aarhus C, Denmark}
\email{dainese@phys.au.dk}

\author[orcid=0000-0001-9825-7418,gname=Kamil,sname=Kalinowski]{Kamil K.\ Kalinowski}
\affiliation{Department of Physics and Astronomy, Aarhus University,
	Ny Munkegade 120, 8000 Aarhus C, Denmark}
\email{kamil@phys.au.dk}

\author[orcid=0000-0001-9985-0643,gname=Caleb,sname=Lammers]{Caleb Lammers}
\affiliation{Department of Astrophysical Sciences, Princeton University,
	4 Ivy Lane, Princeton, NJ 08544, USA}
\email{cl5968@princeton.edu}

\author[orcid=0000-0001-8511-2981,gname=Norio,sname=Narita]{Norio Narita}
\affiliation{Komaba Institute for Science, The University of Tokyo, 3-8-1 Komaba, Meguro, Tokyo 153-8902, Japan}
\affiliation{Astrobiology Center, 2-21-1 Osawa, Mitaka, Tokyo 181-8588, Japan}
\affiliation{Instituto de Astrof\'isica de Canarias (IAC), 38205 La Laguna, Tenerife, Spain}
\email{narita@g.ecc.u-tokyo.ac.jp}

\author[orcid=0000-0003-0987-1593,gname=Enric,sname=Pall\'e]{Enric Pall\'e}
\affiliation{Instituto de Astrof\'isica de Canarias (IAC), 38205 La Laguna, Tenerife, Spain}
\affiliation{Departamento de Astrof\'isica, Universidad de La Laguna (ULL), 38206 La Laguna, Tenerife, Spain}
\email{epalle@iac.es}

\begin{abstract}
	Most of the confirmed circumbinary planets (CBPs) follow orbits with periods
	shorter than a year around stellar binaries with periods shorter than a
	month. The occurrence of CBP systems with longer periods is uncertain,
	largely because of the short-period biases inherent in the transit and
	Doppler techniques that have been used to detect CBPs. Here, we report the
	discovery of a CBP with a period of 6.7~years around an eclipsing binary
	with a period of 124.9 days. The decisive evidence for the planet,
	KIC\,9028474\,b, comes from eclipse-timing variations, supplemented by
	variations in the eclipse depths and a low-significance modulation of the
	binary's systemic radial velocity. Based on Kepler and TESS photometry and
	FIES radial velocities, the planet's mass is $10.48\pm0.27\,M_{\rm Jup}$,
	the orbital eccentricities are $0.1813\pm0.0036$ for the planet and
	$0.82032\pm0.00012$ for the binary, and the angle between the two orbital
	planes is $2.5\pm0.5^\circ$. Among the confirmed CBPs with dynamically
	measured masses around main-sequence binaries, this system has the
	longest-period planetary orbit, the highest planet mass, and the widest and
	most eccentric binary orbit.
\end{abstract}


\section{Introduction}

Circumbinary planets (CBPs) are attractive laboratories for planet formation and
dynamics, but they are difficult to detect \citep[see, e.g.,][]{Martin2018}.
Most of the secure detections have been made with the transit technique,
starting with \cite{Doyle+2011}. This technique favors short-period systems and
coplanar orbits. The Doppler technique is not as biased toward coplanarity
\citep{Triaud+2022}, but still favors short periods. The method of
eclipse-timing variations (ETVs) can also reveal CBPs, and does not require
coplanarity \citep{Goldberg+2023}. Indeed, the related technique of pulsar
timing was used decades ago to discover a body, later shown to have
$\approx$2.5\,$M_{\rm Jup}$, orbiting a pulsar--white-dwarf binary
\citep{Thorsett+1993, Sigurdsson+2003}. Eclipse timing of post-common-envelope
binaries has also produced CBP candidates \citep[e.g.,][]{Lee+2009,Qian+2009},
but the interpretation of the signals is debated \citep{ZorotovicSchreiber2013}.

In principle, astrometry is a useful complement to the other techniques for CBP
detection. It requires neither transits nor eclipses, and its sensitivity
increases with the planet's orbital period as long as the data span a large part
of the period. To this point, adequate astrometry has not been available for
this purpose, but this is expected to change with Data Release 4 (DR4) of the
ESA Gaia Mission, scheduled for 2026
December~2.\footnote{\url{https://www.cosmos.esa.int/web/gaia/data-release-4}}

As part of the BANANA\footnote{Binaries Are Not Always Neatly Aligned} survey
\citep[e.g.,][]{Albrecht+2009, Albrecht+2011, Marcussen+2024}, which measures
spin-orbit angles in double-star systems, we obtained spectra of the binary
KIC~9028474 and analyzed its light curves obtained by Kepler
\citep{Borucki+2010} and TESS \citep{Ricker+2015}. Large eclipse-timing
variations serendipitously revealed a circumbinary planet.

\section{The system and the data}

\citet{Dong+2013} identified KIC~9028474 as a highly eccentric binary system
(eccentricity larger than 0.80) based on the close separation between its
primary and secondary eclipses. KIC~9028474 is an eclipsing pair of solar-type
stars ($M_1=1.21\,M_\odot$ and $M_2=1.04\,M_\odot$) in a 124.9~d orbit of
eccentricity $0.82$. We refer to the more massive star as the primary star,
while noting the peculiarity that the system's eccentricity and inclination
cause the primary eclipses to be shallower than the secondary eclipses.
\cite{Borkovits+2016} first analyzed the system's eclipse-timing variations, and
\cite{KjurkchievaVasilevaAtanasova2017} and \cite{Ozdarcan2023} improved our
knowledge of the system's parameters. Table~\ref{tab:system} gives the
parameters drawn both from the literature and the work described below.

\input{tab_system}

\paragraph{Photometry}

Kepler data are available with 29.4-min sampling from May 2009 to May 2013,
spanning 12 primary and 12 secondary eclipses. One eclipse occurred during a gap
in observing, and two were only partially covered. We used light curves
resulting from simple aperture photometry. The out-of-eclipse point-to-point
scatter is 76\,ppm. Our noise model includes a photon-noise term and a constant
term, to match the scatter:
\[
	\sigma_i^2 = (66.2\,{\rm ppm})^2\,F_i + (37.7\,{\rm ppm})^2,
\]
where $\sigma_i$ is the adopted uncertainty and $F_i$ is the normalized flux.

TESS observed the system in 2024 (Sector 81), catching a pair of eclipses and
extending the baseline to 15 years. We used Quick-Look Pipeline photometry
\citep{Huang+2020a} with 200~s sampling. The noise appears to be dominated by
sources other than photon noise; the scatter does not decrease perceptibly
during eclipses. We adopted a single per-point uncertainty of 3800\,ppm, the
observed scatter.

In total we analyzed 23 eclipses, 21 from Kepler (10 primary, 11 secondary) and
two from TESS, in windows of $\pm9$ hours around each eclipse. The eclipse
light curves are shown in Appendix~\ref{app:additional}.

\begin{figure}
	\centering
	\includegraphics[width=\columnwidth]{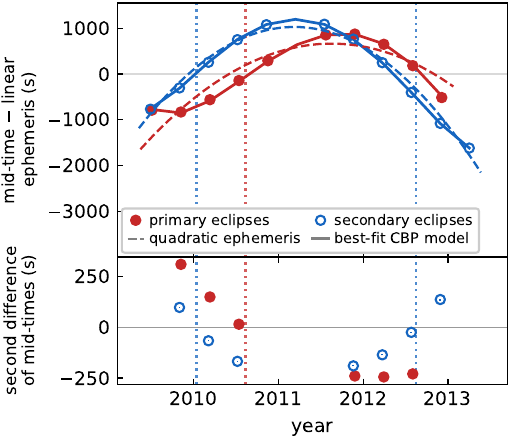}
	\caption{{\bf Kepler ETVs.} {\bf Top}: Deviations of the observed mid-times
		from a linear ephemeris fitted to the 21 Kepler mid-times, with one
		period for both eclipse types (red: primary, blue: secondary).
		Uncertainties are smaller than the symbols. The dashed curves are
		best-fit quadratic ephemerides, which do not describe the data well.
		Solid curves show the best-fit CBP model. {\bf Bottom}: Second
		differences of consecutive mid-times. The vertical dotted lines mark where
		they change sign, i.e., where the curvature of the ETV signal
		reverses.\label{fig:KeplerETVs}}
\end{figure}

\paragraph{Spectroscopy}

We obtained optical high-resolution spectra of the system with the FIES
spectrograph \citep{Telting+2014} at the Nordic Optical Telescope (NOT). Two
out-of-eclipse spectra were obtained in 2014, five in 2017, and another six in
2026. The exposure times were 1380--2400\,s, resulting in a signal-to-noise
ratio per pixel of 10--24 at 5000\,\AA. We computed broadening functions (BFs)
following \citet{Rucinski1992}, using a synthetic PHOENIX template with $T_{\rm
			eff}=6500$\,K, $\log g=4.5$, and $[\mathrm{Fe/H}]=-1.0$ \citep{Husser+2013}. To
measure the radial velocity (RV) of each star, we subtracted the modeled profile
of its companion and fitted a Gaussian to the remaining peak.
Table~\ref{tab:RVs} provides the RVs, with uncertainties arising from the
scatter in the BFs.

\paragraph{Eclipse-timing variations}

We measured the eclipse mid-times by fitting a model in which each eclipse is
allowed to have an independent timing and slow linear trend. The mid-time is the
shift with the lowest chi-squared. Its uncertainty is the shift that raises
chi-squared by one, after scaling each eclipse to a reduced chi-squared of one. These
uncertainties are between 1 and 4\,s for the Kepler eclipses and 16 and 45\,s
for the two TESS eclipses. These values are purely statistical and do not
	include potential timing noise from sources which might lead to subtle flux
	changes on timescales of a few hours. Times and uncertainties are listed with the ephemeris
in Table~\ref{tab:midtimes} in the appendix.

\section{The case for a circumbinary planet}

Before laying out the CBP model in detail, here we present relatively simple
arguments for why the ETVs are caused by a circumbinary body, and how the ETV
characteristics are related to the system parameters.

\paragraph{Ruling out apsidal motion and the R{\o}mer effect}

Figure~\ref{fig:KeplerETVs} shows the deviations between the Kepler mid-eclipse
times and the predictions of a linear ephemeris, assuming the primary and
secondary eclipse periods are identical. The ETVs have peak-to-peak amplitudes
of up to $\approx2700$~s, about three orders of magnitude larger than
the per-eclipse uncertainties.

The timing deviations of primary and secondary eclipses are correlated, whereas
apsidal motion would produce anticorrelated variations. Thus, apsidal motion can
be excluded as the largest contributor to the ETVs. This suggests a third body
as the likely source.

A distant companion shifts eclipse times in two ways. It moves the binary around
the system barycenter, which delays or advances the arrival of the eclipse
signal (the light-travel-time or R{\o}mer effect), and it dynamically perturbs
the inner orbit. Could the signal we see be purely due to the R{\o}mer effect?
Given the 2700~s peak-to-peak amplitude and the 3.37\,yr recurrence
time (see below), one would require a third body of at least
$4.1\,M_\odot$.\footnote{For a pure R{\o}mer effect, half the peak-to-peak
	amplitude times the speed of light is the projected semi-major axis of the
	binary's orbit around the barycenter. With the stellar masses of
	Table~\ref{tab:system} and the 3.37\,yr recurrence time as orbital period, the
	mass function of the outer orbit then gives the minimum mass of the third body.}
Such a massive star would outshine the two components in the binary, and can be
excluded. Alternatively, it could be a stellar-mass black hole. However, such a
dark object would still cause swings of $\sim$25\,km\,s$^{-1}$ in the binary's
systemic radial velocity, which are excluded by our FIES observations. A pure
R{\o}mer delay would also shift primary and secondary eclipses by the same
amount, whereas the observed ETVs of the two eclipse types differ in amplitude
and shape (Figure~\ref{fig:KeplerETVs}). Thus, by process of elimination, the
ETVs are caused by the changing gravity field of the tertiary companion as it
orbits the system barycenter.\footnote{For the companion properties derived
	below, the maximum R{\o}mer effect is only about 10\,s. The light-travel time
	across the binary orbit shifts the primary and secondary eclipses by another 11
	and 6\,s, in opposite directions. Our model includes both.}

\paragraph{The outer period and eccentricity from the ETV pattern}

The circumbinary companion's perturbing potential is dominated by the quadrupole
term
\begin{equation}
	\Phi_{\rm quad} = -\frac{G m_3 r^2}{r_{\rm out}^3} P_2(\cos\Theta),\; P_2(u) = \tfrac{1}{2}(3u^2-1),
\end{equation}
where $G$ is the gravitational constant, $m_3$ is the mass of the companion, $r$
is the separation of the two binary components, $r_{\rm out}$ is the distance
from the binary to the companion, and $\Theta$ the angle between the line
connecting the two binary components and the line from their barycenter to the
companion ($u=\cos\Theta$). During one orbit of the inner pair, $\Theta$ runs
once through a full circle. Since $P_2(\cos\Theta)$ is an even function, the
perturbation is unchanged when the direction to the companion is reversed. If
the outer orbit is circular, $r_{\rm out}$ is also the same half an orbit later,
and we expect ETVs to repeat with period $P_{\rm out}/2$, half of the outer
orbit's period. For an eccentric outer orbit this symmetry is broken, as we
discuss below.

\begin{figure}
	\centering
	\includegraphics[width=\columnwidth]{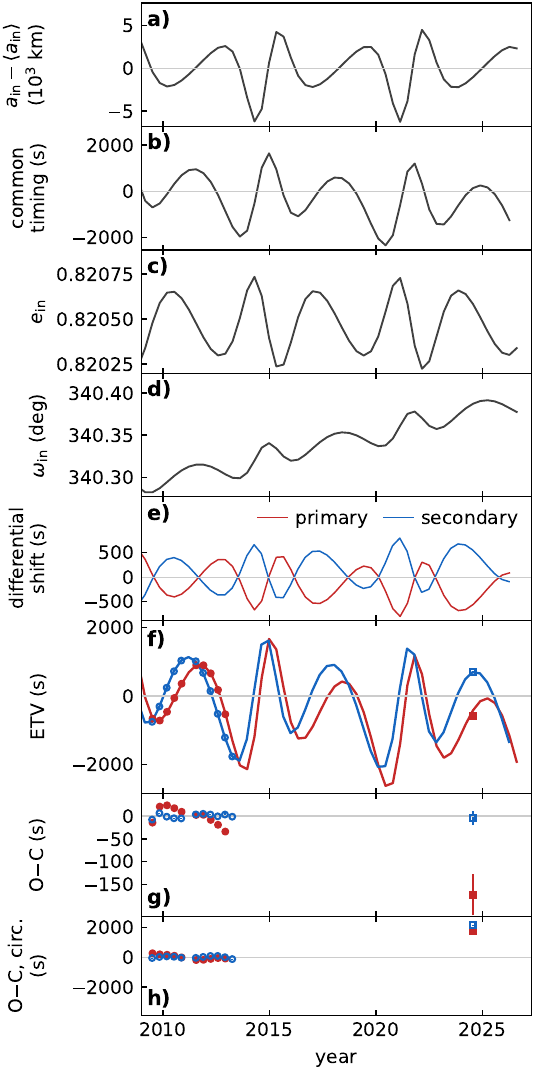}
	\caption{{\bf The companion modulates the orbital elements of the binary.} The curves are calculated
		from the best-fit CBP model.
		\textbf{a)} Inner binary's osculating semi-major axis relative to its mean.
		\textbf{b)} Time shifts common to both eclipse types, caused by the
		modulation of the orbital period, $e_{\rm in}$ and $\omega_{\rm in}$.
		\textbf{c)} Osculating eccentricity.
		\textbf{d)} Argument of periastron.
		\textbf{e)} Time shifts that differ between
		primary and secondary eclipses, caused by
		the effects of modulations in $e_{\rm in}$ and $\omega_{\rm in}$.
		\textbf{f)} Full eclipse-timing model,
		compared with Kepler data (circles) and TESS data (squares).
		\textbf{g)} Residuals about this model.
		\textbf{h)} Residuals about the best-fit model
		in which the outer orbit is constrained
		to be circular (note the wider plot
		range). \label{fig:elements}}
\end{figure}

If the Kepler observations cover about one repetition of the ETV pattern,
roughly half an outer orbit, the parameters of the outer body could be well
constrained. This appears to be the case. Figure~\ref{fig:KeplerETVs} shows that
an {\it ad hoc} quadratic model does not fit, and that both the primary and
secondary ETVs change from positive to negative curvature. This suggests the
ETVs trace a large portion of a full oscillation, rather than a small fraction
of a cycle. The data suggest the recurrence time is $\approx$~3.37\,yr and,
therefore, the outer period is $\approx$~6.7\,yr.

Figure~\ref{fig:KeplerETVs} shows two other patterns in the ETV curves. They are
not quite symmetric around their maxima, and the primary curve seems to lag the
secondary curve. The asymmetry can be explained by a nonzero eccentricity of the
outer body. For an object on an eccentric orbit, the opposite points of the
orbit occur at different distances and are associated with different orbital
speeds, leading to unequal $\Phi_{\rm quad}$ values and breaking the symmetry
between the rising and falling portions of the ETV signals.

\begin{figure*}
	\centering
	\includegraphics[width=0.48\textwidth]{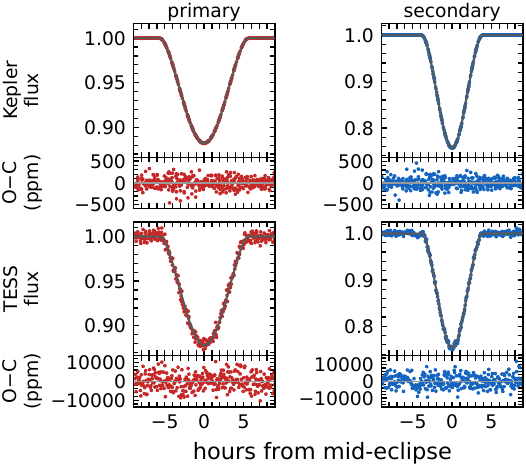}\hfill
	\includegraphics[width=0.48\textwidth]{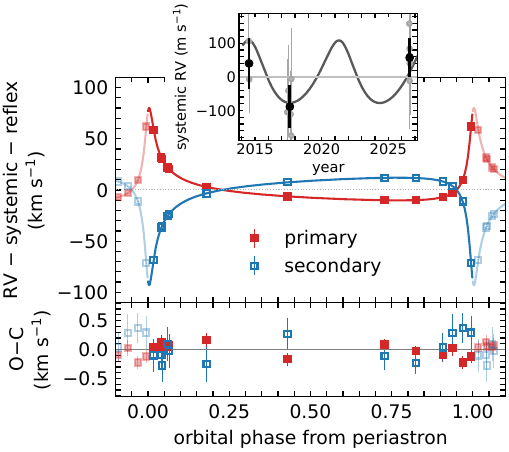}
	\caption{{\bf Best-fit CBP model.}
			{\bf Left}: Stacked eclipse light curves and residuals, after
		correcting for ETVs. The top panels show Kepler data and the bottom
		panels show TESS data. {\bf Right}: RVs as a function of the orbital
		phase of the inner binary, with phase zero at periastron. The
		systemic velocity of each star, the linear drift and the
		planet-induced motion of the binary's barycenter were subtracted
		before plotting. Error bars include the factor of 3.5 applied in the
		fit; faded symbols and curves repeat data and model beyond phases 0
		and 1. The lower panel shows the residuals. The inset panel shows
		the radial velocity of the barycenter of the inner pair: single
		epochs (gray), weighted averages per observing season (black), and
		the model (curve).\label{fig:bestfit-model}}
\end{figure*}

The lag arises from the companion-induced modulation of the inner binary's
eccentricity and argument of periastron ($e_{\rm in}$, $\omega_{\rm in}$; see
Figure~\ref{fig:elements}). For $e_{\rm in} = 0.82$ and $\omega_{\rm in} \approx
	340^\circ$, the values derived below, the secondary eclipse occurs just before
periastron passage and the primary eclipse occurs shortly afterward. When the
companion drives up $e_{\rm in}$, the periastron passage is tighter and is swept
through faster, shifting the times of both eclipses closer to the time of
periastron. Meanwhile, $\omega_{\rm in}$ is modulated by two effects: an
oscillation similar to that in $e_{\rm in}$, and a long-term steady advance
(apsidal motion).\footnote{The apsidal motion in this case is dominated by the
	perturbation from the third body. Tidal and rotational contributions are
	negligible, and in our best-fit model, relativistic precession amounts to only
	$\sim6$~percent of the total advance.} Altering $\omega_{\rm in}$ moves both
conjunctions along the orbit by the same angle, but by different amounts of time
due to Kepler's second law. Since the stars are farther apart at primary
conjunctions than secondary conjunctions, an angular shift of the line of
apsides has a greater effect on the timing of the primary eclipse. Since these
time shifts operate differently on primary and secondary eclipses, they
introduce a phase lag between the ETV curves. Furthermore, reversing the outer
orbit's direction  reverses the phasing of the modulation and with it the sign
of the lag. Although it is not obvious, the observed sign establishes that the
companion orbits in the same direction as the binary.

Finally, the 2700\,s peak-to-peak amplitude of the ETVs sets the mass of the
companion. Relative to the binary's own potential, $-G(M_1+M_2)/r$, $\Phi_{\rm
		quad}$ is of order $[m_3/(M_1+M_2)]\,(r/r_{\rm out})^3$. Kepler's third law for
both orbits turns this into $\epsilon \approx [m_3/(M_1+M_2+m_3)]\,(P_{\rm in}/P_{\rm
	out})^2$. This fractional perturbation of the inner orbit acts over a time of
order $P_{\rm out}$ and so shifts the eclipses by about $\epsilon P_{\rm out}$,
which for the observed amplitude gives $m_3$ of order $10\,M_{\rm Jup}$.

\paragraph{Information in the eclipse depths}
There may also be information in changes of the eclipse depths. The depths are
governed by the eclipse impact parameters,
\begin{equation}
	\begin{aligned}
		b_{\rm pri} & = \frac{a_{\rm in}\cos i_{\rm in}}{R_1}\,
		\frac{1-e_{\rm in}^2}{1+e_{\rm in}\sin\omega_{\rm in}}, \\[6pt]
		b_{\rm sec} & = \frac{a_{\rm in}\cos i_{\rm in}}{R_2}\,
		\frac{1-e_{\rm in}^2}{1-e_{\rm in}\sin\omega_{\rm in}}.
	\end{aligned}
	\label{eq:impact}
\end{equation}
Here $a_{\rm in}$ is the semi-major axis of the binary, and $R_1$ and $R_2$ are
the radii of the two stars. Two mechanisms can change these impact parameters.
First, if the outer orbit is inclined against the inner orbit, the torque of the
companion causes the binary orbit to precess. The precession changes the
line-of-sight inclination $i_{\rm in}$, and with it the eclipse depths. Second,
changes in $e_{\rm in}$ and $\omega_{\rm in}$ change the distance between the
two stars at the eclipse phases, again leading to changes in impact parameters
and eclipse depths. As we will see below, both effects are relevant in this
system.

\section{Determination of planet parameters} \label{sec:numbers}

We now describe the quantitative determination of the circumbinary planet
parameters, using three-body numerical integrations. We used a
``photodynamical'' approach, i.e., we fitted the light curves directly rather
than fitting eclipse properties that were extracted from the light curves. We
fitted the RVs alongside the light curves.

We used the \textsc{rebound} package with the IAS15 integrator
\citep{ReinLiu2012, ReinSpiegel2015}, including the first-order post-Newtonian
corrections of all three bodies (gr\_full from the \textsc{reboundx} library;
\citealt{Tamayo+2020}). The model has a total of 29 parameters, identified with
asterisks in Table~\ref{tab:system}. Eclipse profiles were computed using the
\textsc{occultquad} routine of \citet{MandelAgol2002} and \citet{Eastman+2013}
assuming a quadratic limb-darkening law, and averaged over the exposure times
of 29.4\,min (Kepler) and 200\,s (TESS). Light-curve detrending was performed
for each model evaluation by optimizing a linear function of time covering each
eclipse window. The effects of variable blending in the images were accounted
for by a dilution parameter $s$ for TESS and for each Kepler quarter: in each
eclipse window the model flux is $(a+bt)\,[1-s\,(1-F)]$, where $F$ is the flux
of the binary normalized to unity outside the eclipses and $a+bt$ the linear
function of time.

We optimized the parameter values by using the Nelder--Mead simplex algorithm to
minimize the $\chi^2$ statistic. We found that the uncertainties of the RVs and
the Kepler photometry were too small to properly represent the mismatch between
the data and the best-fit model. For the Kepler photometry, the root-mean-square
residual is larger inside the eclipses (1.7 times the noise model) than outside
(1.3 times). We therefore scaled the uncertainties such that the best-fit model
has $\chi^2=N_{\rm dof}$ before determining parameter uncertainties. The Kepler
flux uncertainties were increased by a factor of 1.58 and the RV uncertainties
by a factor of 3.5. For TESS we adopted a single flux uncertainty of 3800\,ppm,
0.93 times the median of the formal uncertainties.

Table~\ref{tab:system} gives the parameter values and uncertainties, based on a
Markov Chain Monte Carlo analysis with \textsc{emcee}
\citep{ForemanMackey+2013}. We used uniform priors on the fitted parameters,
with the limb darkening restricted to physical values, and ran 64 walkers,
started close to the best fit, for 50{,}000 steps. We discarded the first 4000
steps, slightly more than twice the longest integrated autocorrelation time; the
remaining 46{,}000 steps span 26 autocorrelation times. Figure~\ref{fig:mcmc} in
Appendix~\ref{app:additional} shows the posteriors.

The residuals of the primary-eclipse mid-times about the best-fit model
are correlated over years (panel~g of Figure~\ref{fig:elements}). With the
measurement uncertainties removed in quadrature, the rms of the Kepler mid-time
residuals is 17.5\,s for primary and 4.2\,s for secondary eclipses. To account
for these in the final parameter uncertainties we carried out two additional
fits. First we refitted the data, allowing the model in each eclipse window a
time shift, restricted by Gaussians with widths of 17.5 and 4.2 s, for primary
and secondary respectively. For these fits, we again scaled the Kepler and RV
uncertainties so that $\chi^2=N_{\rm dof}$. In a first fit, the shifts were
independent of each other. In a second fit, the shifts of eclipses of the same
type were in addition correlated, with a correlation of $\exp[-(\Delta
		t/164\,{\rm d})^2/2]$ for two eclipses $\Delta t$ apart. 164\,d is the
autocorrelation time of the residuals of the ten Kepler primary eclipses. The
second fit changed the best-fitting values more than the first, and we used it.
For each parameter, we added its uncertainty in quadrature to the change of its
best-fitting value from a fit without shifts. Where the result exceeded the
uncertainty of the original posterior by more than 10\,percent, we adopted it
(daggers in Table~\ref{tab:system}). Figure~\ref{fig:bestfit-model} shows the
best-fit model compared with the stacked eclipse light curves and the RVs.

\section{Results} \label{sec:results}

\begin{figure}
	\centering
	\includegraphics[width=\columnwidth]{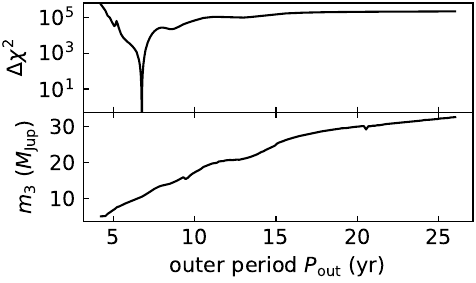}
	\caption{{\bf The outer period is determined uniquely.}
			{\bf Top}: $\chi^2$ of the fits with $P_{\rm out}$ held fixed, relative to the lowest of them.
			{\bf Bottom}: Best-fit companion mass for each choice of $P_{\rm out}$.\label{fig:profile}}
\end{figure}

\paragraph{Planet parameters}
We find a companion mass of $m_3 = 10.48 \pm 0.27\,M_{\rm Jup}$, an
outer period $P_{\rm out} = 2464.8 \pm 32.3$\,d (6.748\,yr), an outer
eccentricity $e_{\rm out} = 0.1813 \pm 0.0036$, and a small but nonzero
mutual inclination between the two orbital planes, $i_{\rm
			mut}=2.48\pm0.53^\circ$.  This configuration is dynamically stable,
since the periastron distance of the outer orbit is $1.9$ times the critical
value ($2.0$\,au) of the stability criterion of \citet{MardlingAarseth2001}. An
integration of the best-fit configuration over 1\,Myr confirms this: the
planet's eccentricity stays between 0.16 and 0.21, and copies of the system
started with the planet displaced by $10^{-8}$\,au do not drift apart. In a run
without the relativistic corrections, the chaos indicator MEGNO \citep[Mean
	Exponential Growth factor of Nearby Orbits;][]{CincottaSimo2000} settles at
2.01, the value expected for a regular, quasi-periodic orbit.

\paragraph{Robustness checks}
To check on the uniqueness of the solution for the CBP orbital period, we
optimized a series of models in which $P_{\rm out}$ was held fixed at values
between 4 and 26\,yr. We refitted all the other parameters, except the
radial-velocity semi-amplitudes, and held the planet's orbit coplanar with the
binary orbit. Figure~\ref{fig:profile} shows the results. The $\chi^2$ value
comes within 100 of the minimum only for $P_{\rm out} = 6.68$--$6.82$\,yr. Thus,
the solution with $P_{\rm out} \approx	6.7$~yr is singled out.

\paragraph{Radial velocities}
Averaged per observing season, the systemic velocities follow the
high--low--high pattern of the predicted barycenter motion (see the inset panel
of Figure~\ref{fig:bestfit-model}). Although the RV coverage is not complete
enough for the observed pattern to be diagnostic of a third body, the pattern is
consistent with the CBP solution.

\begin{figure}[t]
	\centering
	\includegraphics[width=\columnwidth]{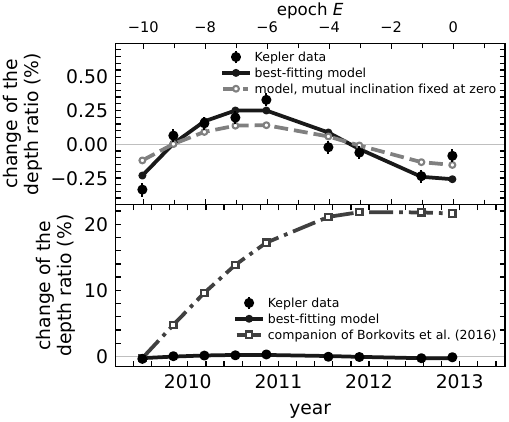}
	\caption{{\bf Ratio of the primary to the secondary eclipse depth} in the Kepler data,
		relative to its mean, against time and epoch $E$ (upper axis). The depth is the flux deficit
		at mid-eclipse relative to the local out-of-eclipse flux. It is measured in the same way on the
		data and on the model light curves, and within one Kepler quarter the ratio does not depend on
		the dilution parameters.
			{\bf Top}: Data (filled circles), the best-fitting model (solid line), and the model with the
		mutual inclination fixed at zero (dashed line).
			{\bf Bottom}: The same data and best-fitting model, with the prediction (dash-dotted line) for the companion of
		\citet{Borkovits+2016}, added to our inner orbit at epoch $-10$.
		\label{fig:shape}}
\end{figure}

\paragraph{Eclipse depths}
The eclipse depths also carry information about the mutual inclination. Because
we allowed a free dilution parameter for each Kepler quarter
(Section~\ref{sec:numbers}), we cannot compare the depths themselves from
quarter to quarter, only the ratio of the primary to the secondary eclipse depth
within each quarter in which both eclipses were observed. This ratio, shown in
Figure~\ref{fig:shape}, does not depend on the dilution parameter. It does
depend on the inclination of the binary orbit and on $e_{\rm in}$ and
$\omega_{\rm in}$. With $e_{\rm in}=0.82$ and $\omega_{\rm in}=340^\circ$, the
impact parameters are $b_{\rm pri}R_1=1.63\,R_\odot$ and $b_{\rm
			sec}R_2=0.93\,R_\odot$ (Equation~\ref{eq:impact}). Both scale with $\cos i_{\rm
			in}$, but the more grazing primary eclipse responds more strongly, so a change
of $i_{\rm in}$ changes the ratio of the two eclipse depths. In the best-fitting
model (solid line), the inclination of the binary orbit changes in addition to
$e_{\rm in}$ and $\omega_{\rm in}$. Measured on their own, the depth ratios
favor this model over the model with the mutual inclination fixed at zero
(dashed line) by about 20 in $\chi^2$. The two TESS eclipses of 2024 cannot
distinguish between the exactly coplanar and nearly coplanar ($2.5^\circ$)
models, because the fitted TESS dilution parameter and light ratio can absorb
changes of their depths since the Kepler years. A lot of information is lost
this way, but even so the fit with a free mutual inclination of $2.5^\circ$ is
preferred over the exactly coplanar model by $\Delta\chi^2 = 32$ over all data,
for two additional free parameters. However, most of this $\Delta\chi^2$ arises
in just two primary eclipses, at epochs $-10$ and $-6$. Refitting both
models without the eclipse at epoch $-10$ reduces $\Delta\chi^2$ to 20, without
the eclipse at epoch $-6$ to 13, and without both eclipses to 11, which still
favors the free mutual inclination at about $3\sigma$. Yet without eclipses $-6$
and $-10$ only the lines of nodes of the two orbits on the sky differ
($\Omega_{\rm out}=0.46\pm0.14^\circ$), whereas the two orbital inclinations
relative to the line of sight agree within their uncertainty ($i_{\rm
			out}-i_{\rm in}=-0.3\pm0.7^\circ$). For this reason we treat the nonzero mutual
inclination with some caution, and the measured value might also be regarded as
an upper limit.

\paragraph{A possible fourth body?}
While our model successfully describes most of the data, it is not a perfect
fit. Panel g of Figure~\ref{fig:elements} shows that the residuals of the
primary-eclipse times show a slow, coherent structure. The structure resembles
an additional apsidal drift of the inner orbit, which our model cannot
reproduce, as the apsidal rate of the three-body integration follows from the
masses and the orbits.

Starspots are an unlikely origin of this additional apsidal drift. For a spot to
modify the primary-eclipse times by $\sim 30$~s, one would require a flux
distortion of about 400\,ppm. Modulations of this size cannot be excluded within
single Kepler quarters, but the residual timing structure is coherent over
years, while spot patterns should change from orbit to orbit. Invoking a fourth
body in the system might explain the residuals, but as we have not found any
other evidence for such a body, we did not investigate further.

\paragraph{Comparison with a prior eclipse-timing solution}
KIC~9028474 was part of an eclipse-timing study of the Kepler multiple-star
candidates by \citet{Borkovits+2016}. They fitted analytic formulae for the
combined light-travel and dynamical delays to the Kepler mid-times, finding a
companion with $P_{\rm out} = 3378 \pm 94$\,d, $m_3 = 0.04 \pm 0.02$\,$M_\odot$
(about $42 \pm 21\,M_{\rm Jup}$), an outer orbit with $e_{\rm out} = 0.09 \pm
	0.02$, and a mutual inclination of $50.6 \pm 0.9^\circ$. Their solution
describes the different apparent amplitudes and shapes of the primary and
secondary ETV curves with a nearly circular, strongly inclined outer orbit. In
our model, these features are caused by the eccentricity of the outer orbit. The
decisive evidence in favor of our model is the 2024 TESS eclipse timings,
obtained after the original study. A circular orbit fitted to the Kepler data
alone misses the TESS mid-times by about an hour, and a circular orbit refitted
to the Kepler and TESS data together still leaves residuals a factor of
15 above those of the eccentric solution, as shown in panel h of
Figure~\ref{fig:elements}. Freeing the stellar masses and the orientation of the
outer orbit does not help: the fit then raises the primary mass by 20 percent,
in conflict with the radial velocities. Thus the outer orbit is eccentric and
close to coplanar, not circular and strongly tilted.

\begin{figure*}[t]
	\centering
	\includegraphics[width=\textwidth]{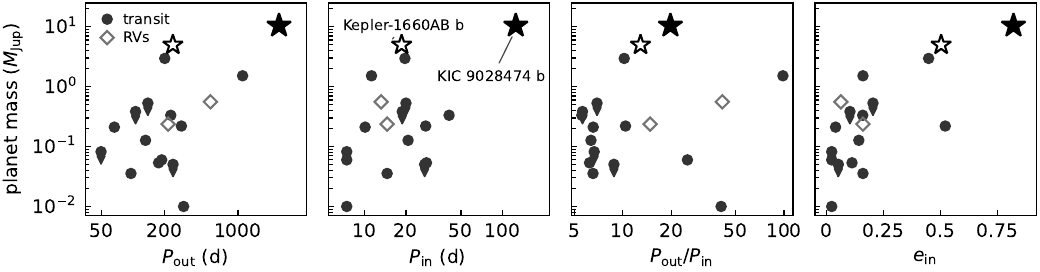}
	\caption{{\bf KIC~9028474\,b among the circumbinary planets.} Confirmed CBPs
	with dynamically measured masses around main-sequence binaries, found
	through transits (circles), radial velocities (diamonds), or eclipse timing
	(stars: open for Kepler-1660AB\,b, filled for this work). Gray arrows mark
	upper mass limits (Kepler-38\,b, Kepler-47\,b [\citealt{Orosz+2019}],
	Kepler-64\,b and Kepler-453\,b). {\bf From left to right}: Planet mass
	against orbital period, against the orbital period of the host binary,
	against period ratio, and against the eccentricity of the binary orbit.
	Planet masses, periods and discovery methods were taken from the NASA
	Exoplanet Archive \citep{Christiansen+2025}, the binary periods from the
	papers that report the planets, and the binary eccentricities from
	exoplanet.eu \citep{Schneider+2011}. \label{fig:context}}
\end{figure*}

A mutual inclination of $50^\circ$ is independently excluded by the eclipses
themselves. With the companion of \citet{Borkovits+2016} added to our inner
orbit, nodal precession would change the inclination of the binary orbit by
$0.17^\circ$ within the Kepler years and by $0.84^\circ$ by 2024. The ratio of
the primary to the secondary eclipse depth would then grow by 22 percent within
the Kepler years (bottom panel of Figure~\ref{fig:shape}). By 2024 the primary
eclipse would last 13.7 instead of the observed 11.1 hours, a change that no
dilution parameter or light ratio can absorb (see the TESS primary eclipse in
Figure~\ref{fig:bestfit-model}). The same type of reasoning settled the
architecture of Kepler-1660AB\,b, where an orbit inferred to be misaligned from
eclipse timing was ruled out by the absence of eclipse depth variations
\citep{Goldberg+2023}.

\section{KIC~9028474 among its peers}

Figure~\ref{fig:context} shows key parameters of KIC~9028474\,b and other
confirmed CBPs around main-sequence binaries with dynamically measured masses
from transits, eclipse timing, or radial velocities. In this sample, which by
construction only includes companions with masses below the deuterium-burning
limit, KIC~9028474\,b is the most massive planet. It has twice the mass of the
runner-up, Kepler-1660AB\,b ($4.99\pm0.23\,M_{\rm Jup}$,
\citealt{Goldberg+2023}), which was also discovered with ETVs.\footnote{An
unconfirmed eclipse-timing candidate, first reported by \citet{Borkovits+2016},
orbits the eclipsing pair of the 2+2 quadruple KIC~7177553 \citep{Lehmann+2016},
with a mass of $8.1\pm0.6\,M_{\rm Jup}$ \citep{Borkovits+2025}.} It also has the
longest orbital period, 2.2 times that of Kepler-1647\,b \citep{Kostov+2016}.
Furthermore, KIC~9028474 is the widest CBP-hosting binary in this sample, with
$P_{\rm in}=124.9$\,d. The previous record holder is Kepler-16 at 41.1\,d
(\citealt{Doyle+2011}; second panel of Figure~\ref{fig:context}). On the other
hand, KIC~9028474's period ratio $P_{\rm out}/P_{\rm in} = 19.7$ is
unremarkable, as is its nearly coplanar orbit. Its planet-to-binary mass ratio,
0.0044, is the highest in the sample, twice that of Kepler-1660AB\,b. Its binary
orbit is also the most eccentric among the CBP systems in
Figure~\ref{fig:context}, with $e_{\rm in}=0.82$, against 0.52 for the
runner-up, Kepler-34 \citep{Welsh+2012}.

Two directly imaged companions of young main-sequence binaries have reported
masses similar to that of KIC~9028474\,b, but those masses are based on
evolutionary models, and their orbits are far wider: b~Centauri\,(AB)b
($10.9\pm1.6\,M_{\rm Jup}$ at $\approx560$\,au, \citealt{Janson+2021}) and
HD~106906\,b ($11\pm2\,M_{\rm Jup}$, \citealt{Bailey+2014}; $\approx740$\,au,
\citealt{NguyenDeRosaKalas2021}). The sequence of circumbinary companions with
dynamical masses continues above the limit with BEBOP-4\,(AB)\,b, a brown dwarf
with $m\sin i = 20.9\pm1.3\,M_{\rm Jup}$ on a 5.0-yr orbit around a 72-day
eclipsing binary \citep{Triaud+2025}. Further brown-dwarf-mass candidates from
eclipse timing remain unverified \citep{Borkovits+2025}, and the eclipse-timing
triples of the Kepler field have tertiaries mostly of stellar mass
\citep{Borkovits+2016}. With two objects between 5 and 25\,$M_{\rm Jup}$, the
data cannot distinguish a continuous mass function from a gap, although such
massive circumbinary companions appear rare \citep{Martin+2019}.

The dynamical ETV method has enhanced sensitivity to wide binary orbits,
because the dynamical delay scales as $P_{\rm in}^2/P_{\rm out}$. The
astrometric method has a preference for massive planets on distant orbits,
like KIC~9028474\,b. However, it will be unlikely that we will be able to
detect KIC~9028474\,b in the upcoming Gaia DR4 release. We simulated the
Gaia measurements of the system using \textsc{gaiamock}
\citep{ElBadry+2024} and following the detection threshold of
	\citet{LammersWinn2026}. The planet would remain undetected in a DR4 blind
	search, because of its large distance (623\,pc) and its orbital period
	longer than the DR4 time series (5.5~yr). Would we be able to
	confirm our specific CBP model? For this we  fixed the binary and
	planetary orbits at our solution, except for their common line of nodes on
	the sky. Our simulations find that the DR4 data would confirm this specific planet at the
	$3\sigma$ level in 29 percent of the simulated data sets. Gaia's Data
	Release 5 (DR5), spanning the entire 10.5\,yr of the mission, would confirm
	it in 96 percent.

\section{Summary}

The eclipse times of the eccentric, 125-day double-lined eclipsing binary
KIC~9028474 wander by up to three quarters of an hour. The eclipse depths also
vary subtly, as does the binary's systemic RV. We carried out an analysis of the
system, using Kepler and TESS light curves and FIES radial velocities. Almost
all of the anomalies are well explained as consequences of dynamical
perturbations of the binary orbit by a circumbinary giant planet with a mass of
$10.5\,M_{\rm Jup}$ on a 6.7-year orbit with an eccentricity of 0.18 and a
mutual inclination of $2.5^\circ$. The model's only failure is that the timing
residuals display a coherent structure on the order of tens of seconds. A fourth
body might be responsible, though this possibility needs further investigation.

KIC~9028474\,b is the most massive and the longest-period circumbinary planet
with a dynamically measured mass around a main-sequence binary. It also orbits
the widest and most eccentric (0.82) binary known to host such a planet, and is
located in a region of parameter space that is hard to reach for transit and RV
surveys, and more accessible to eclipse timing and astrometry.

Gaia DR4 might detect a number of systems like KIC~9028474, if they are
	closer to the Sun and host a massive planet with a period of about 4\,yr or
	shorter, and DR5 should find more (see also the forecasts by
	\citealt{SahlmannTriaudMartin2015} and \citealt{BaycroftTriaudSahlmann2026}).
	With the DR4 and DR5 data we may learn
a lot more about the occurrence, formation, and dynamics of these fascinating
systems.

\begin{acknowledgments}
	We are grateful to the NOT staff for their excellent support during the
	observations. We thank Jerome de Leon for his prompt help during the
	preparation of this work. S.H.A., S.D., and K.K.K.\ acknowledge support from
	the Independent Research Fund Denmark (Danmarks Frie Forskningsfond) through
	grant 5241-00072B. This work is partly supported by the National Natural
	Science Foundation of China (grant No. 12633012). This work is partly
	supported by JSPS KAKENHI Grant Numbers JP24H00017, JP25K24620, JP26H01402,
	and JP26K00755. Based on observations made with the Nordic Optical
	Telescope, owned in collaboration by the University of Turku and Aarhus
	University, and operated jointly by Aarhus University, the University of
	Turku and the University of Oslo, representing Denmark, Finland and Norway,
	the University of Iceland and Stockholm University at the Observatorio del
	Roque de los Muchachos, La Palma, Spain, of the Instituto de Astrofisica de
	Canarias, under programs 49-404, 55-011, 73-410, and 73-804. This paper
	includes data collected by the Kepler and TESS
	missions, obtained from the MAST data archive at the
	Space Telescope Science Institute, which is operated by the Association of
	Universities for Research in Astronomy, Inc., under NASA contract
	NAS5-26555. Funding for the Kepler and TESS missions is provided by the NASA
	Science Mission Directorate. All the Kepler and TESS data used in this paper
	can be found in MAST: the Kepler long-cadence light curves \citep{MAST_KeplerLC},
	\dataset[10.17909/T9488N]{https://doi.org/10.17909/T9488N}, and
	the TESS light curve of Sector 81 \citep{Albrecht2026},
	\dataset[10.17909/64rc-8392]{https://doi.org/10.17909/64rc-8392}.
	The comparison sample of Figure~\ref{fig:context} uses the Planetary Systems
	table of the NASA Exoplanet Archive \citep{NEA_PS},
	\dataset[10.26133/NEA12]{https://doi.org/10.26133/NEA12}.
\end{acknowledgments}

\facilities{Kepler, TESS, NOT(FIES)}

\software{\textsc{rebound} \citep{ReinLiu2012, ReinSpiegel2015},
	\textsc{reboundx} \citep{Tamayo+2020},  \textsc{numpy} \citep{Harris+2020},
	\textsc{scipy} \citep{Virtanen+2020}, \textsc{matplotlib}
	\citep{Hunter2007}, \textsc{astropy} \citep{AstropyCollaboration2022},
	\textsc{emcee} \citep{ForemanMackey+2013}, \textsc{occultquad}
	\citep{MandelAgol2002, Eastman+2013}, \textsc{corner}
	\citep{ForemanMackey2016}, \textsc{gaiamock} \citep{ElBadry+2024}}

\appendix
\restartappendixnumbering

\section{Additional figures and tables} \label{app:additional}

\begin{figure*}[p!]
	\centering
	\includegraphics[width=\textwidth,height=0.93\textheight,keepaspectratio]{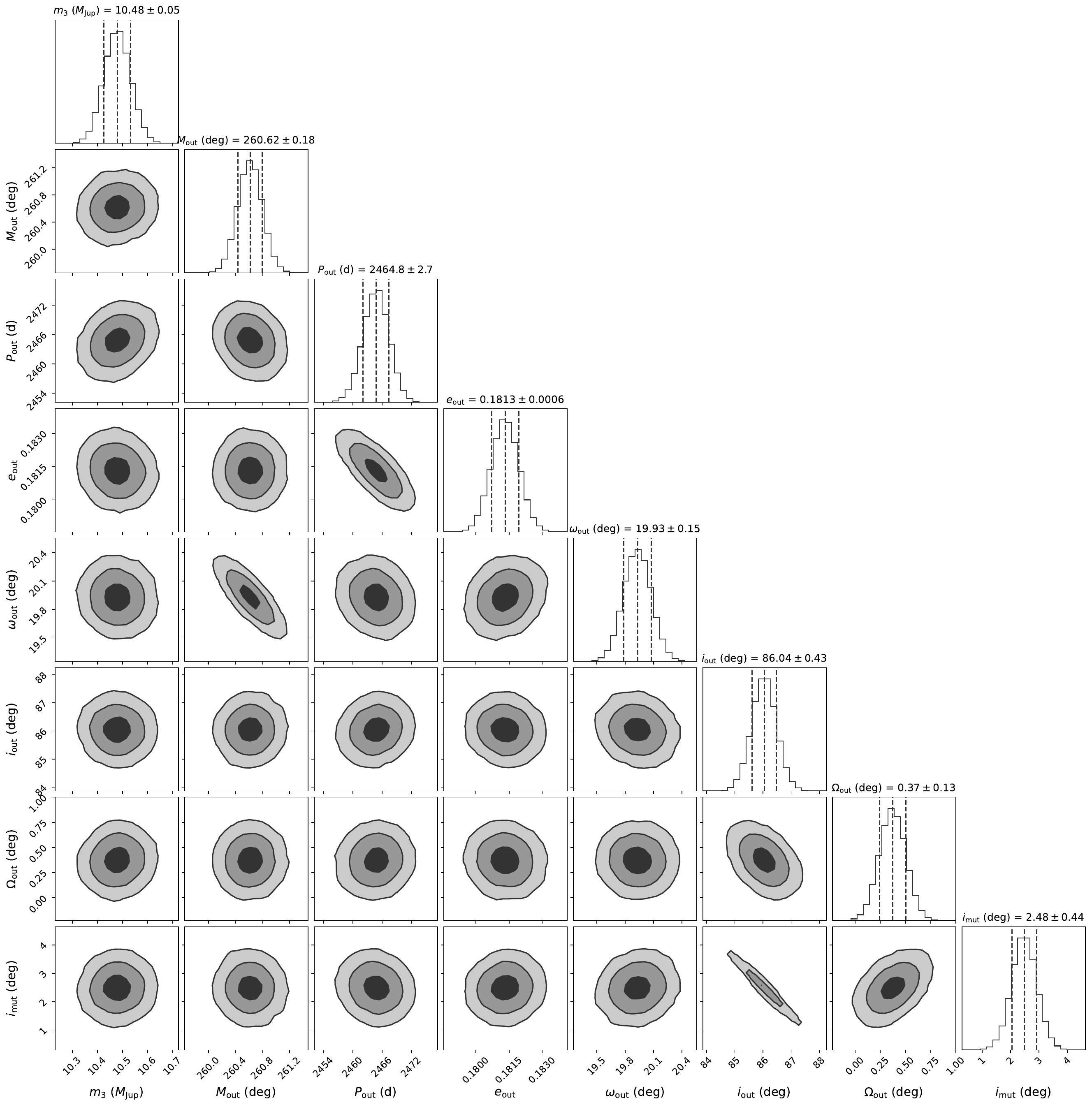}
	\caption{{\bf Posterior distributions of the planet parameters.}
		Mass, mean anomaly at $T_0$, orbital period, eccentricity, argument of
		periastron, inclination and line of nodes of the outer orbit, and the
		mutual inclination between the two orbital planes. Contours are the
		1$\sigma$, 2$\sigma$ and 3$\sigma$ levels of the two-dimensional
		distributions, enclosing 39, 86 and 99 percent of the samples.
		The uncertainties marked with daggers in Table~\ref{tab:system}
			are larger than the widths of these distributions, because they also
			account for the correlated timing residuals (Section~\ref{sec:numbers}).
		\label{fig:mcmc}}
\end{figure*}

\begin{figure*}[p!]
	\centering
	\includegraphics[width=\textwidth,height=0.86\textheight,keepaspectratio]{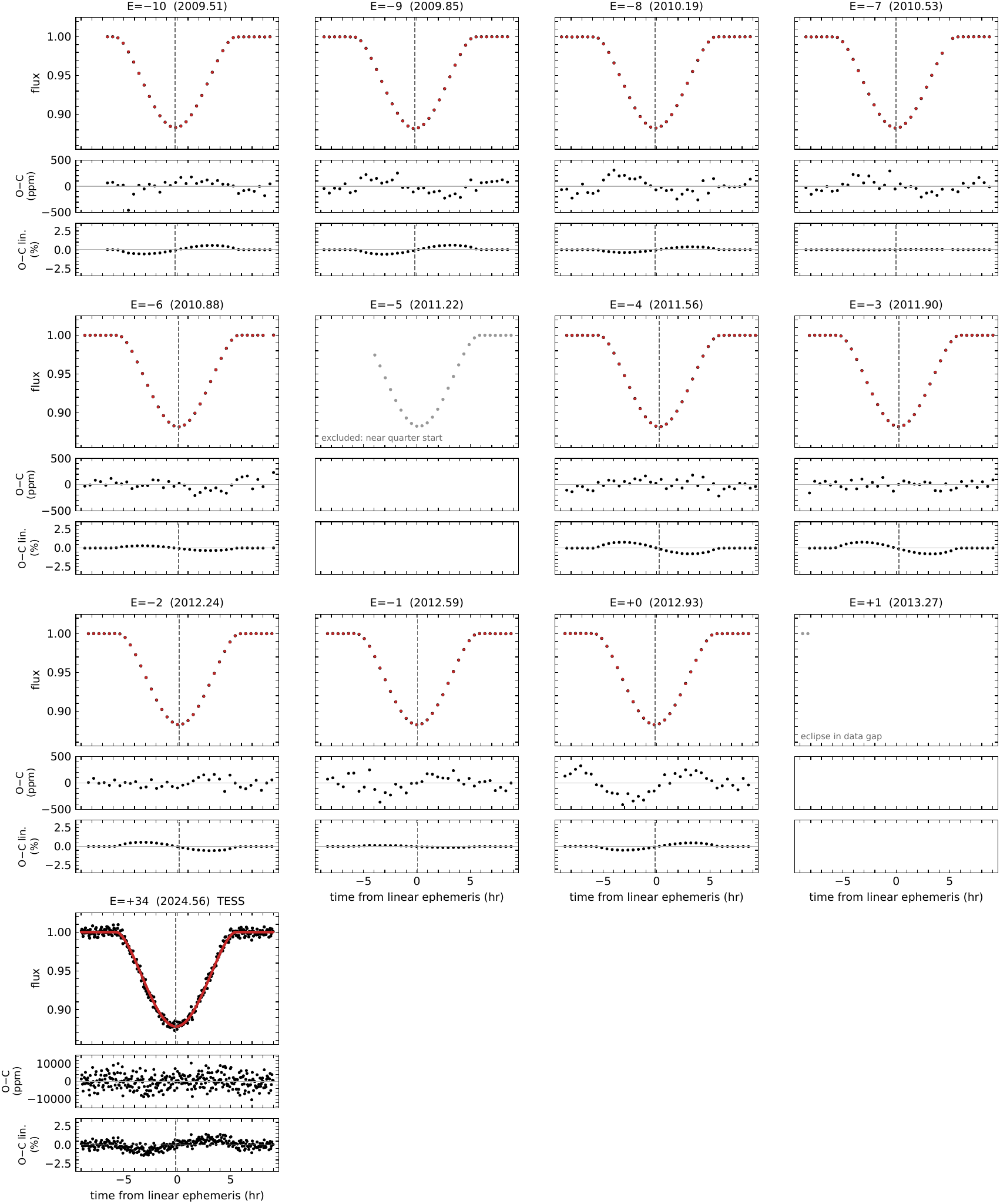}
	\caption{{\bf All primary eclipses.} Each eclipse, labeled by epoch $E$ and
		fractional year, is shown in a block of three panels on a common time
		axis centered on the mid-time predicted by the joint linear ephemeris
		(Table~\ref{tab:midtimes}). {\bf Top}: Normalized flux (black) and
		best-fit CBP model (red). {\bf Middle}: Residuals between the data and
		the best-fit CBP model. {\bf Bottom}: Residuals between the data and the
		best-fit CBP model shifted to the joint linear ephemeris. The excursions
		are due to eclipse-timing variations. The dashed vertical lines in the
		top and bottom panels mark the measured mid-eclipse times; offsets from
		zero are eclipse-timing variations. The last block shows the 2024 TESS
		eclipse. Gray points are plotted for data that were excluded from the
		analysis (near a quarter start, or in a data gap).
		\label{fig:gallery_pri}}
\end{figure*}

\begin{figure*}[p!]
	\centering
	\includegraphics[width=\textwidth,height=0.93\textheight,keepaspectratio]{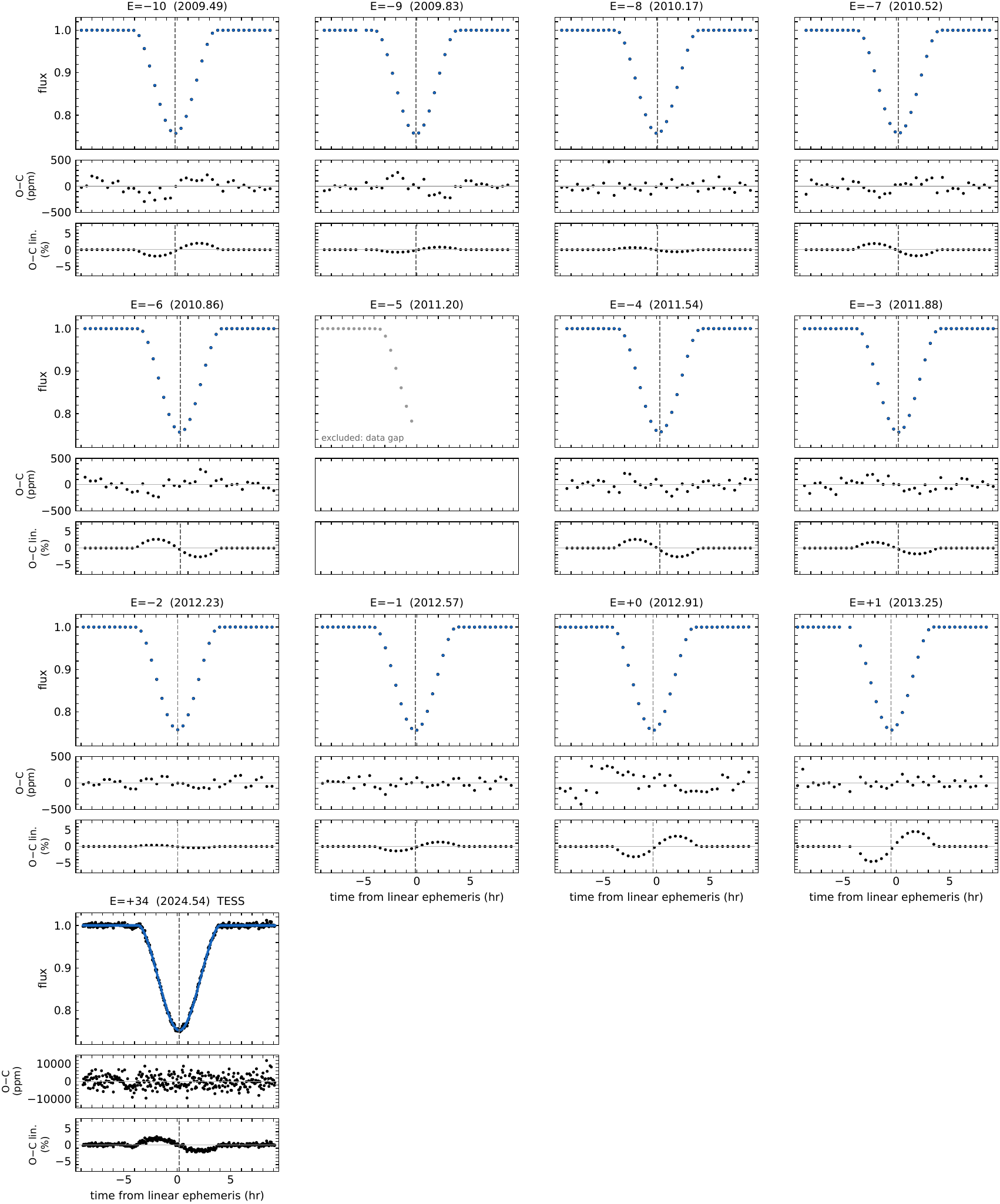}
	\caption{{\bf All secondary eclipses.}
		Same as in Figure~\ref{fig:gallery_pri}, but for the secondary
		eclipses, and with the CBP model in blue.
		\label{fig:gallery_sec}}
\end{figure*}

Figures~\ref{fig:gallery_pri} and \ref{fig:gallery_sec} show every primary and
secondary eclipse in the Kepler time span, together with the 2024 TESS eclipses;
the events excluded from the analysis are shown in gray. Figure~\ref{fig:mcmc}
shows the posterior distributions of the planet parameters. Table~\ref{tab:RVs}
lists the FIES radial velocities, and Table~\ref{tab:midtimes} the measured
eclipse mid-times and their deviations from the joint linear ephemeris.

\input{tab_rvs}
\input{tab_midtimes}

\bibliography{kic9028474_refs}{}
\bibliographystyle{aasjournalv7.1}

\end{document}

%% file: tab_system.tex
\begin{table}
	\centering
	\scriptsize
	\setlength{\tabcolsep}{0.8pt}
	\caption{System parameters \label{tab:system}}
	\begin{tabular}{llr@{\hspace{1pt}$\pm$\hspace{1pt}}l}
	\hline\hline
	Parameter & Symbol & \multicolumn{2}{c}{Value} \\
	\hline
	\multicolumn{4}{c}{\textit{Identifiers, astrometry, photometry}} \\
	Identifiers & \multicolumn{3}{l}{KIC 9028474, KOI 3510, TIC 270856764} \\
	Gaia DR3 & \multicolumn{3}{l}{2127962476185944576} \\
	Coordinates (J2000) & \multicolumn{3}{l}{$19^{\rm h}36^{\rm m}38\fs29$, $+45^{\circ}22'33\farcs3$} \\
	Parallax & $\varpi$\,(mas) & $1.5908$ & $0.0093$ \\
	Distance & $d$\,(pc) & $623$ & $4$ \\
	Magnitudes & $G$, $J$\,(mag) & \multicolumn{2}{l}{12.28, 11.19} \\
	Spectral types & & \multicolumn{2}{l}{F9\,V $+$ G1\,V} \\
	\multicolumn{4}{c}{\textit{Inner binary}} \\
	Orbital period$^\star$ & $P_{\rm in}$\,(d) & $124.937057$ & $0.000060^\dagger$ \\
	Primary conjunction$^\star$ & $T_0$\,(BJD$_{\rm TDB}$) & $2456266.36596$ & $0.00023^\dagger$ \\
	Eccentricity$^\star$ & $e_{\rm in}$ & $0.82032$ & $0.00012$ \\
	Argument of periastron$^\star$ & $\omega_{\rm in}$\,(deg) & $340.30$ & $0.07$ \\
	Inclination$^\star$ & $i_{\rm in}$\,(deg) & $88.498$ & $0.011$ \\
	Scaled radius, pri$^\star$ & $R_1/a_{\rm in}$ & $0.01116$ & $0.00001$ \\
	Scaled radius, sec$^\star$ & $R_2/a_{\rm in}$ & $0.00772$ & $0.00004$ \\
	Ratio of the radii & $R_2/R_1$ & $0.6919$ & $0.0031$ \\
	Light ratio, Kepler$^\star$ & $(L_2/L_1)_{\rm Kep}$ & $0.4373$ & $0.0046$ \\
	Light ratio, TESS$^\star$ & $(L_2/L_1)_{\rm TESS}$ & $0.4418$ & $0.0057$ \\
	Limb darkening, pri$^\star$ & $(2u_1+u_2)_{\rm Kep}$ & $1.020$ & $0.028$ \\
	Limb darkening, sec$^\star$ & $(2u_1+u_2)_{\rm Kep}$ & $1.098$ & $0.032$ \\
	Limb darkening, pri$^\star$ & $(2u_1+u_2)_{\rm TESS}$ & $0.76$ & $0.14$ \\
	Limb darkening, sec$^\star$ & $(2u_1+u_2)_{\rm TESS}$ & $0.83$ & $0.15$ \\
	RV amplitude, pri$^\star$ & $K_1$\,(km\,s$^{-1}$) & $45.116$ & $0.079$ \\
	RV amplitude, sec$^\star$ & $K_2$\,(km\,s$^{-1}$) & $52.451$ & $0.145$ \\
	Systemic RV at $T_0$, pri$^\star$ & $\gamma_1$\,(km\,s$^{-1}$) & $-18.316$ & $0.054$ \\
	Systemic RV at $T_0$, sec$^\star$ & $\gamma_2$\,(km\,s$^{-1}$) & $-18.045$ & $0.097$ \\
	Linear RV drift$^\star$ & $\dot\gamma$\,(m\,s$^{-1}$\,yr$^{-1}$) & $6.6$ & $7.6$ \\
	Mass, pri & $M_1$\,($M_\odot$) & $1.210$ & $0.007$ \\
	Mass, sec & $M_2$\,($M_\odot$) & $1.041$ & $0.005$ \\
	Radius, pri & $R_1$\,($R_\odot$) & $1.539$ & $0.003$ \\
	Radius, sec & $R_2$\,($R_\odot$) & $1.065$ & $0.005$ \\
	Surface gravity, pri & $\log g_1$\,(cgs) & $4.147$ & $0.002$ \\
	Surface gravity, sec & $\log g_2$\,(cgs) & $4.401$ & $0.004$ \\
	Semi-major axis & $a_{\rm in}$\,(au) & $0.6410$ & $0.0010$ \\
	Effective temp., pri & $T_{\rm eff,1}$\,(K) & $6030$ & $100$ \\
	Effective temp., sec & $T_{\rm eff,2}$\,(K) & $5890$ & $100$ \\
	Metallicity & [Fe/H]\,(dex) & $+0.12$ & $0.08$ \\
	Age & (Gyr) & $4.9$ & $0.5$ \\
	Rotation, pri & $v\sin i_1$\,(km\,s$^{-1}$) & $3.6$ & $1.0$ \\
	Rotation, sec & $v\sin i_2$\,(km\,s$^{-1}$) & $2.4$ & $1.0$ \\
	\multicolumn{4}{c}{\textit{Circumbinary planet}} \\
	Mass$^\star$ & $m_3$\,($M_{\rm Jup}$) & $10.48$ & $0.27^\dagger$ \\
	Mean anomaly at $T_0$$^\star$ & $M_{\rm out}$\,(deg) & $260.62$ & $0.49^\dagger$ \\
	Orbital period$^\star$ & $P_{\rm out}$\,(d) & $2464.8$ & $32.3^\dagger$ \\
	Orbital period & $P_{\rm out}$\,(yr) & $6.748$ & $0.088^\dagger$ \\
	Eccentricity$^\star$ & $e_{\rm out}$ & $0.1813$ & $0.0036^\dagger$ \\
	Argument of periastron$^\star$ & $\omega_{\rm out}$\,(deg) & $19.93$ & $0.81^\dagger$ \\
	Inclination$^\star$ & $i_{\rm out}$\,(deg) & $86.04$ & $0.51^\dagger$ \\
	Line of nodes$^\star$ & $\Omega_{\rm out}$\,(deg) & $0.37$ & $0.25^\dagger$ \\
	Semi-major axis & $a_{\rm out}$\,(au) & $4.6873$ & $0.0410^\dagger$ \\
	Mutual inclination & $i_{\rm mut}$\,(deg) & $2.48$ & $0.53^\dagger$ \\
	Barycenter RV amplitude & $K_{\rm bary}$\,(m\,s$^{-1}$) & $92.8$ & $1.9^\dagger$ \\
	\hline
	\end{tabular}
	\par\smallskip
	\begin{minipage}{\columnwidth}\footnotesize
	\textsc{Note}---Identifiers, coordinates,
    parallax, and magnitudes are from SIMBAD \citep{Wenger+2000}, Gaia DR3 \citep{GaiaCollaboration+2016, GaiaCollaboration+2023},
    and 2MASS \citep{Skrutskie+2006}. Spectral types are from \cite{Ozdarcan2023}. Quoted parameters are based on the 16/50/84th percentiles of the posteriors.
    Asterisks mark fitted parameters, daggers uncertainties enlarged for correlated timing residuals (Section~\ref{sec:numbers}).
    The distance is the inverse of the parallax corrected for the zero point of $-0.0145$\,mas \citep{Lindegren+2021}.
    All orbital elements are osculating Jacobi elements at BJD$_{\rm TDB}$ 2456266.3714, near the $E=0$ primary eclipse.
    $T_{\rm eff}$, [Fe/H], and age
	are based on  BaSTI isochrones \citep{Hidalgo+2018} fitted to the masses, radii, light ratios, and parallax.
    Their uncertainties are conservative adopted values (formal: 60\,K, 0.06\,dex, and 0.3\,Gyr).
    $v\sin i$ values are from the FIES spectra.
	\end{minipage}
\end{table}

%% file: tab_rvs.tex
\begin{deluxetable}{lccrclrcl}
	\tablecaption{FIES radial velocities \label{tab:RVs}}
	\tablehead{\colhead{Mid-exposure} & \colhead{Date} & \colhead{Exp.\ time} &
		\multicolumn{3}{c}{RV primary} & \multicolumn{3}{c}{RV secondary} \\
		\colhead{(BJD$_{\rm TDB}$)} & \colhead{(UT)} & \colhead{(s)} & \multicolumn{3}{c}{(km\,s$^{-1}$)} & \multicolumn{3}{c}{(km\,s$^{-1}$)}}
	\startdata
	2456852.57753 & 2014-07-14 & 2400 & $-28.259$ & $\pm$ & $0.024$ & $-6.233$ & $\pm$ & $0.068$ \\
	2456864.58782 & 2014-07-26 & 2400 & $-28.224$ & $\pm$ & $0.023$ & $-6.519$ & $\pm$ & $0.056$ \\
	2457908.50495 & 2017-06-04 & 1380 & $-15.340$ & $\pm$ & $0.037$ & $-21.668$ & $\pm$ & $0.087$ \\
	2457939.60910 & 2017-07-05 & 1500 & $-24.851$ & $\pm$ & $0.033$ & $-10.472$ & $\pm$ & $0.075$ \\
	2457999.41086 & 2017-09-02 & 1500 & $-25.515$ & $\pm$ & $0.035$ & $-9.850$ & $\pm$ & $0.069$ \\
	2458016.35766 & 2017-09-19 & 1500 & $+11.708$ & $\pm$ & $0.034$ & $-53.257$ & $\pm$ & $0.080$ \\
	2458018.41434 & 2017-09-21 & 1500 & $+3.648$ & $\pm$ & $0.035$ & $-43.522$ & $\pm$ & $0.085$ \\
	2461251.49835 & 2026-07-29 & 1500 & $-21.256$ & $\pm$ & $0.040$ & $-14.092$ & $\pm$ & $0.095$ \\
	2461255.40098 & 2026-08-02 & 1500 & $-8.459$ & $\pm$ & $0.032$ & $-29.174$ & $\pm$ & $0.074$ \\
	2461258.60472 & 2026-08-06 & 1500 & $+43.561$ & $\pm$ & $0.033$ & $-89.599$ & $\pm$ & $0.083$ \\
	2461261.41213 & 2026-08-08 & 1500 & $+40.322$ & $\pm$ & $0.032$ & $-86.042$ & $\pm$ & $0.080$ \\
	2461264.41111 & 2026-08-11 & 1500 & $+13.428$ & $\pm$ & $0.033$ & $-54.661$ & $\pm$ & $0.068$ \\
	2461267.38513 & 2026-08-14 & 1500 & $+1.994$ & $\pm$ & $0.035$ & $-41.375$ & $\pm$ & $0.085$ \\
	\enddata
\end{deluxetable}

%% file: tab_midtimes.tex
\begin{deluxetable}{llrccr}
	\tablecaption{Measured eclipse mid-times \label{tab:midtimes}}
	\tablehead{\colhead{Instrument} & \colhead{Type} & \colhead{Epoch} & \colhead{Mid-eclipse time} & \colhead{Uncertainty} & \colhead{O$-$C} \\
		\colhead{} & \colhead{} & \colhead{} & \colhead{(BJD$_{\rm TDB}$)} & \colhead{(s)} & \colhead{(s)}}
	\startdata
	Kepler & secondary & $-10$ & 2455010.67040 & 1.1 & $-750$ \\
	Kepler & primary & $-10$ & 2455016.98624 & 3.9 & $-658$ \\
	Kepler & secondary & $-9$ & 2455135.61283 & 1.0 & $-302$ \\
	Kepler & primary & $-9$ & 2455141.92283 & 2.7 & $-713$ \\
	Kepler & secondary & $-8$ & 2455260.55639 & 1.4 & $+244$ \\
	Kepler & primary & $-8$ & 2455266.86302 & 3.0 & $-459$ \\
	Kepler & secondary & $-7$ & 2455385.49918 & 1.0 & $+724$ \\
	Kepler & primary & $-7$ & 2455391.80492 & 3.0 & $-56$ \\
	Kepler & secondary & $-6$ & 2455510.44003 & 1.2 & $+1035$ \\
	Kepler & primary & $-6$ & 2455516.74700 & 3.1 & $+362$ \\
	Kepler & secondary & $-4$ & 2455760.31429 & 1.1 & $+1016$ \\
	Kepler & primary & $-4$ & 2455766.62767 & 3.3 & $+898$ \\
	Kepler & secondary & $-3$ & 2455885.24760 & 1.0 & $+677$ \\
	Kepler & primary & $-3$ & 2455891.56494 & 2.6 & $+900$ \\
	Kepler & secondary & $-2$ & 2456010.17871 & 0.9 & $+148$ \\
	Kepler & primary & $-2$ & 2456016.49944 & 2.6 & $+663$ \\
	Kepler & secondary & $-1$ & 2456135.10825 & 1.1 & $-517$ \\
	Kepler & primary & $-1$ & 2456141.43110 & 3.9 & $+181$ \\
	Kepler & secondary & $0$ & 2456260.03750 & 2.4 & $-1208$ \\
	Kepler & primary & $0$ & 2456266.36010 & 3.8 & $-530$ \\
	Kepler & secondary & $+1$ & 2456384.96833 & 1.1 & $-1762$ \\
	TESS & secondary & $+34$ & 2460507.92563 & 15.6 & $+693$ \\
	TESS & primary & $+34$ & 2460514.22556 & 45.4 & $-588$
	\enddata
	\tablecomments{O$-$C is the measured mid-time minus the joint linear ephemeris, with one shared period of 124.937239\,d and one epoch per eclipse type, 2456266.36624 (primary) and 2456260.05148 (secondary, both BJD$_{\rm TDB}$), determined by unweighted least squares from all 23 mid-times.}
\end{deluxetable}